\documentclass[preprint]{elsarticle}
\usepackage{graphicx}

\begin{document}

\begin{frontmatter}
\title{Accuracy Test for Link Prediction in terms of Similarity Index: The Case of WS and BA Models}

\author[a]{Min-Woo Ahn}

\author[a,b]{Woo-Sung Jung\corref{cor}}
\cortext[cor]{Corresponding author}
\ead{wsjung@postech.ac.kr}

\address[a]{Department of Physics, Pohang University of Science and Technology, Pohang, 790-784, Republic of Korea}
\address[b]{Department of Industrial and Management Engineering, Pohang University of Science and Technology, Pohang, 790-784, Republic of Korea}

\begin{abstract}
Link prediction is a technique that uses the topological information in a given network to infer the missing links in it.
Since past research on link prediction has primarily focused on enhancing performance for given empirical systems, negligible attention has been devoted to link prediction with regard to network models.
In this paper, we thus apply link prediction to two network models: The Watts-Strogatz (WS) model and Barab{\'a}si-Albert (BA) model.
We attempt to gain a better understanding of the relation between accuracy and each network parameter (mean degree, the number of nodes and the rewiring probability in the WS model) through network models.
Six similarity indices are used, with precision and area under the ROC curve (AUC) value as the accuracy metrics.
We observe a positive correlation between mean degree and accuracy, and size independence of the AUC value.
\end{abstract}


\journal{Physica A}

\end{frontmatter}

\section{Introduction}

Complex networks have lately attracted considerable attention as a tool for understanding the structure of relation among agents in the system.
Several systems, including social interaction network \cite{mobile,col2,col3,email}, the World-Wide Web \cite{www,www2,www3}, biological networks \cite{bio1,bio2,bio3,bio4} and economic databases \cite{economic}, have been studied to understand the structure of systems through the data.
To obtain more accurate results from network analysis, we need to obtain more complete data.
However, collecting data from the real world is difficult, such that the analyzed data is usually incomplete, i.e. the connection with the real world might not be represented in the data.
We should check all unconnected node pairs in a network to distinguish them in a missing link (a certain relation that exists in the real world, but is not represented in the data) from a nonexistent link (when node pairs relate to each other neither in the real world nor the data).
Link prediction can be used to render data collection more efficient and the data thus collected more complete.
Link prediction corresponds to a process for the calculation of the likelihood of the existence of a link based on observed links, and helps discover and restore the missing links.
If the predictions are sufficiently accurate, we can efficiently identify missed connections by rechecking the data with this information.

Link prediction has been studied to find missing connections in an observed network \cite{HSM,local,lrw,colfilter,colfilter2,citation,wikipedia,block}.
Zhou \emph{et al.} suggested a similarity index from the resource allocation process, and compared the results with other local similarity indices \cite{local}.
Liu and L\"u suggested a few methods based on the local random walk process \cite{lrw}.
Similarity between two nodes was defined as the probability that one node can reach the other.
This index can consider both global and local structures depending on the number of random walk steps.
Clauset \emph{et al.} employed a hierarchical random graph to understand the structure of real-world networks and infer missing links \cite{HSM}.
A hierarchical random graph is constructed from a set of nodes and the connection probabilities among the members of the set, represented by a dendrogram.
Monte-Carlo simulation is used to obtain dendrograms from a given real-world network,
and predicted missing links using the average connection probability calculated from the ensembles of dendrograms.
Guimer{\`a} \emph{et al.} developed a block model \cite{block} that infers missing links as well as spurious connections (when nodes are connected in the data but no corresponding relation obtains in the real system).
L\"u and Zhou have also surveyed previous link prediction methodologies \cite{linkreview}.
Link prediction using collaborative filtering, which has been applied to recommendation systems, can also be included into this class \cite{colfilter, colfilter2}.
Missing link prediction in citation networks \cite{citation} and in Wikipedia \cite{wikipedia} have also been performed.

Link prediction has also been applied to forecast emerging links from the network picture at a given point in time \cite{linkcol, collaboration, human}.
Liben-Nowell and Kleinberg applied link prediction to a collaboration network to predict future links \cite{linkcol}, and employed and compared many kind of similarity indices.
Newman discovered that the probability of a new connection in the collaboration network is proportional to the number of common neighbors \cite{collaboration}.
D. Wang \emph{et al.} used a similarity index and social network information based on human mobility pattern \cite{human}.

In spite of the above-mentioned research, many questions related to link prediction remain unanswered.
Previous studies usually applied link prediction to real-world data.
Although empirical tests are important to understand a given complex system and the methodology, they provide a limited perspective.
As network models can provide network samples under varied conditions,
understanding the methodology using basic network models is also important.
In this paper, we focus on the statistical relation between network parameters and the accuracy of methodologies.

This paper is organized as follows.
In section 2, we describe link prediction in the network models.
We employ two basic network models: the Watts-Strogatz (WS) model, and the Barab{\'a}si-Albert (BA) model.
Six indices are employed: Common Neighbor, Adamic-Adar, Resource Allocation, Jaccard, Preferential attachment, and Simrank.
We use two metrics, precision and area under the ROC curve (AUC) value, as an accuracy measures.
The results are detailed in Section 3 and discussed in Section 4.
Finally, we summarize our findings in Section 5.

\section{Methodology}

The application of link prediction to network models consists of three steps: 
1) missing link creation, 2) similarity calculation, and 3) accuracy calculation.
The first step, the ``preliminary step" of link prediction, is to create missing links in a given network.
We randomly select and remove links, which are considered ``missing" in later steps.
The similarity calculation is the actual prediction step, where we form a list from the similarity index to find the missing links.
Accuracy is calculated from the list in the final step.

\subsection{Dataset}

The control parameters provided are limited in real-world networks because they are difficult to control.
For this reason, we employ two network models: the WS model \cite{WS} and the BA model \cite{BA}.
The WS model, proposed by Watts and Strogatz, describes how real-world networks exhibit a small-world effect and high transitivity.
The BA model, proposed by Barab{\'a}si and Albert, provides an insight into how the degree distribution of real-world networks obeys power-law behavior.
Network models provide various conditions and network ensembles for each condition,
because of which we can observe the statistical relation between each network parameter and the accuracy of methodologies.

We control three parameters: The number of nodes $ N $, the mean degree $ \langle k \rangle $, and the rewiring probability $p$ in the WS model.
We set the parameters to $N_{WS}=N_{BA}=1000$, $\langle k \rangle_{WS}=6$ (in the WS model), $\langle k \rangle_{BA}=10$ (in the BA model) and $p=0.1$ as representative of default condition,
and observe the correlation between each parameter and each accuracy metric.
To calculate accuracy, we tested 1000 network ensembles for each network condition.

\subsection{Missing link creation}

To test link prediction in a static network, we applied the missing link creation process \cite{HSM, lrw, local}.
We can randomly divide the links into the training set (remaining link) and the probe set (missing link), where the links in the probe set are treated as the missing links.
Approximately 10\% of the links were removed and 100 independent creation processes were considered for each network ensemble.

\subsection{Similarity calculation}

We employ similarity indices to obtain the topological information of the remaining network.
A similarity index represents the structural proximity of two nodes,
which corresponds to the possibility of a connection between an unconnected node pair.
In other words, if the members of an unconnected node pair are highly similar, we assume that a missed connection exists between them.
For this reason, a node pair with a high similarity value is assigned a high priority when we check unconnected node pairs.
We use the following six indices in this paper.

\begin{enumerate}

\item Common neighbors (CN)

This index was suggested from the growth mechanism of the coauthorship network \cite{collaboration}.
New links appear when they have a neighbors in common,
and the probability of a connection increases with the number of common neighbors.
This index is defined as
\begin{equation}
S_{CN}(x,y) = |\Gamma (x) \cap \Gamma (y)|,
\end{equation}
where $\Gamma(x)$ is the set of neighbors of a node x and $S_{CN}$ is the similarity value.
\\

\item Adamic-Adar index (AA)

This index was suggested to define similarity between people from their personal homepages, defined by Adamic and Adar \cite{AA}.
They assumed that two people have a close relationship when they have common interests, such as favorite music, and hobbies.
They also assume that they have more close relationship when they share unique items.
This uniqueness corresponds to the degree of common neighbors when we apply this index in a network \cite{linkcol}.
This index is defined as
\begin{equation}
S_{AA}(x,y) = \sum_{z \in \Gamma (x) \cap \Gamma (y)} \frac{1}{log(|k_{z}|)},
\end{equation}
where $k_{z}$ is the degree of node z.
\\

\item Resource allocation index (RA)

Consider an unconnected node pair consisting of nodes $x$ and $y$.
Starting from the node $x$, resources are equally allocated to its neighbors.
We can then define an index for the amount of resources that node $y$ receives \cite{local}.
This index is defined as
\begin{equation}
S_{RA}(x,y) = \sum_{z \in \Gamma (x) \cap \Gamma (y)} \frac{1}{k_{z}}.
\end{equation}
\\

\item Jaccard index

This index was suggested by Jaccard (1901) \cite{jaccard, linkcol}, and is defined as
\begin{equation}
S_{J}(x,y) = \frac{| \Gamma (x) \cap \Gamma (y) |}{|\Gamma(x) \cup \Gamma(y)|}.
\end{equation}
\\

\item Preferential attachment index (PA)

This index is related to the growth mechanism of networks \cite{BA,linkcol,colfilter,local}.
The probability that a new link is connected to a node is proportional to the degree of that node, which is called preferential attachment.
Preferential attachment is used in network model to describe power-law degree distribution of web-page networks \cite{BA}.
Newman revealed that preferential attachment also appears in the collaboration networks \cite{collaboration}.
This index is defined as
\begin{equation}
S_{PA}(x,y) = |\Gamma(x)| \times |\Gamma(y)|.
\end{equation}
\\

\item Simrank

Jeh and Widom suggested this index \cite{sim}.
Two nodes are similar when their neighbors have a close relationship.
This index is defined as
\begin{equation}
S_{Sim} (x,y) = \frac{C}{|\Gamma(x)||\Gamma(y)|} \sum_{x' \in \Gamma(x) , y' \in \Gamma(y)} S_{Sim}(x',y'),
\end{equation}
where C is a constant between 0 and 1.
Simrank can be obtained by the recursive calculation of an arbitrary given similarity matrix (each element in the matrix has a value from 0 to 1).
\\

\end{enumerate}

We first calculate similarity for all unconnected node pairs on the remaining network and form our list.
Following the calculation, we sort this list in descending order of similarity and check whether the links in the list are missing link or not.
We calculate accuracy from this result in the next step.

\subsection{Accuracy calculation}

The accuracy metric is employed to quantify the performance of each index.
Two standard metrics are employed: precision \cite{lrw,colfilter} and area under the ROC curve (AUC) value \cite{ROC,HSM}.

When we obtain the top-$L$ links from the list (where $L$ can be an arbitrary value) and $l$ links are missing links (included in the probe set), the precision is defined as $l/L$.
We set $L$ as the number of the deleted links in the missing link creation step.

\begin{table}
\begin{tabular}{| p{2.5cm} | c | c |}
\hline
& Missing links & Nonexistent links \\
\hline
 & & \\
Classified to & True Positive & False Positive \\
missing links & (TP) & (FP) \\
 & & \\
\hline
 & & \\
Classified to & False Negative & True Negative \\
nonexistent links & (FN) & (TN) \\
 & & \\
\hline
\end{tabular}
\caption{The confusion matrix in link prediction.}
\label{table 1}
\end{table}

\begin{figure}
\includegraphics[width=7cm]{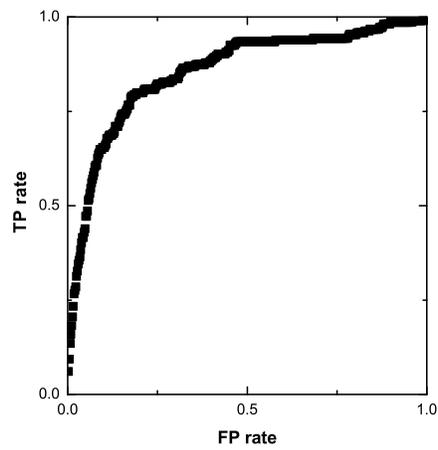}
\caption{An example of a receiver operating characteristic (ROC) curve. We consider the top-$L$ links, and calculate the true positive (TP) rate and the false positive (FP) rate. We obtain the curve by varying $L$. AUC value is the area under the ROC curve.}
\label{fig1}
\end{figure}

The AUC value is defined from the receiver operating characteristic (ROC) curve \cite{ROC}. 
To obtain the ROC curve, we calculate the true positive (TP) rate and the false positive (FP) rate, defined as follows:

\begin{equation}
TP rate = \frac{TP}{TP+FN}, 
FP rate = \frac{FP}{FP+TN},
\end{equation}

where $TP$, $TN$, $FP$ and $FN$ are described in Table. \ref{table 1}.
We treat the top-$L$ links as the missing links, and obtain the ROC curve by calculating the TP rate and the FP rate for all $L$ links, as in FIG. \ref{fig1}.
The AUC value is defined as the area under the ROC curve.

In case of a random classifier, the TP rate and the FP rate are the same for all conditions, and the AUC value is 0.5.
A perfect classifier, should perfectly classify the missing links and the nonexistent links,
because of which the similarity index of the missing links is always higher than that of nonexistent links.
The FP rate is 0 until the TP rate reaches 1, and thus the maximum AUC value is 1.
In sum, a good classifier has an AUC value close to 1, and a poor classifier has one less than 0.5, which is worse than a random classifier.

We calculate two accuracy metrics for each ensemble and each deletion process, and then average them for each condition.

\section{Results}
\begin{figure*}
\includegraphics[width=13cm]{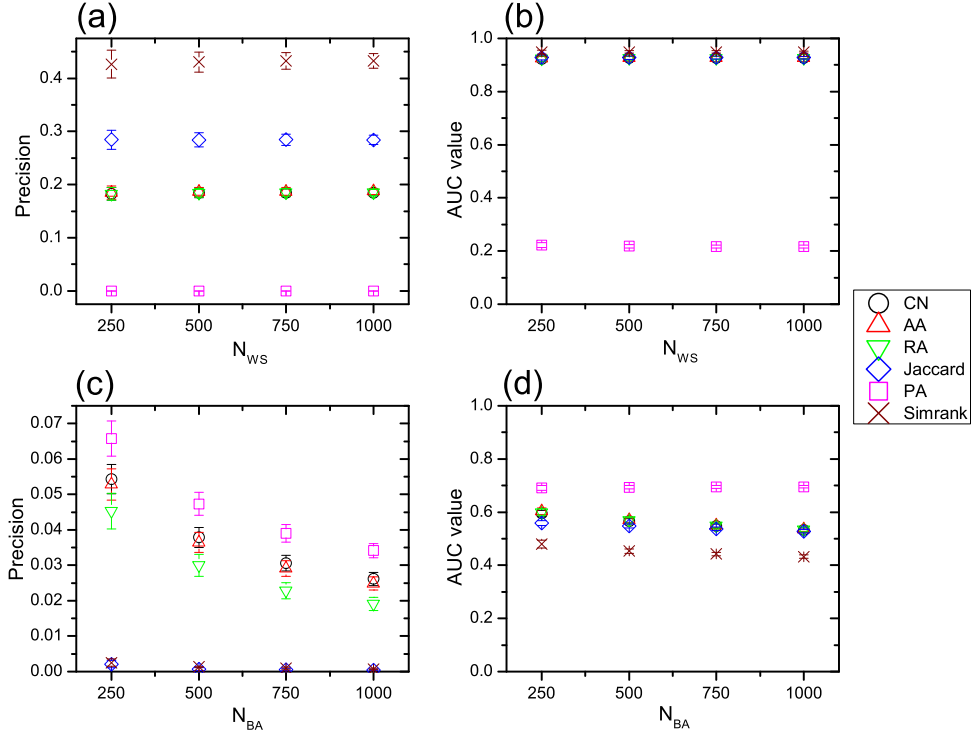}
\caption{Correlation between (a) precision and $N_{WS}$, (b) AUC value and $N_{WS}$, (c) precision and $N_{BA}$ and (d) AUC value and $N_{BA}$.
We set C as 0.8 when we calculate $S_{Sim}$.}
\label{fig2}
\end{figure*}
We first observe the relationship between accuracy and $N$ (Fig. \ref{fig2}).
There is no correlation between accuracy and $N_{WS}$ in the WS model, but precision and $N_{BA}$ exhibit a negative correlation in the BA model.
There is no significant difference between the number of nodes in the two cases with regard to the AUC value.
Both accuracy metrics are independent of $N$ in the WS model because its structure does not depend on the number of nodes.

\begin{figure*}
\includegraphics[width=13cm]{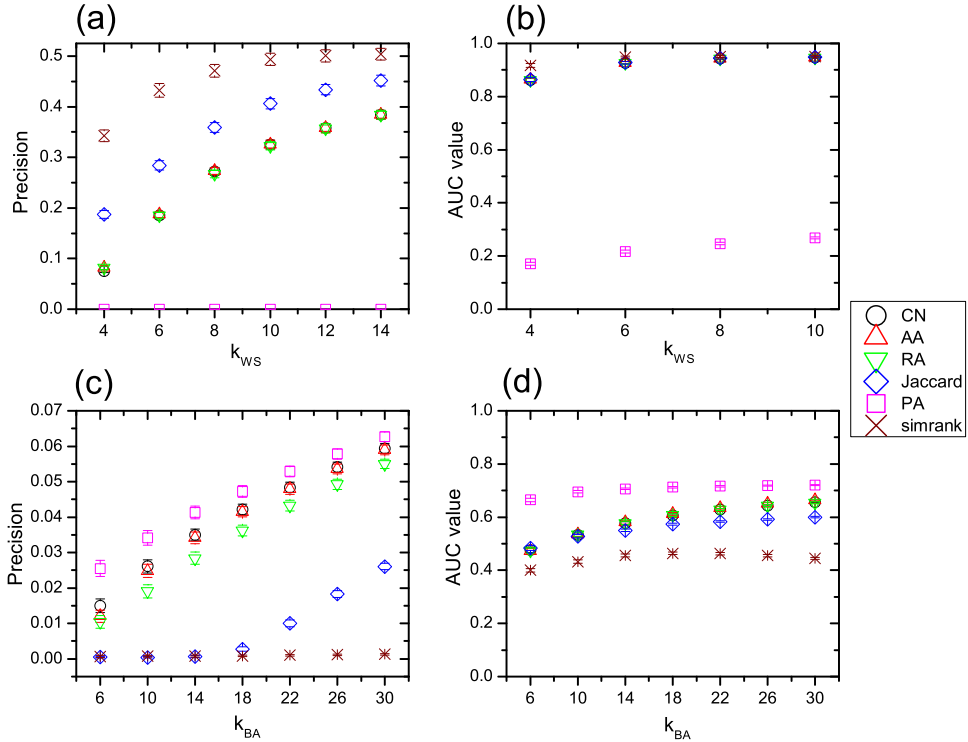}
\caption{Correlation between (a) precision and $ \langle k \rangle_{WS} $, (b) AUC value and $ \langle k \rangle_{WS} $, (c) precision and $ \langle k \rangle_{BA} $ and  (d) AUC value and $ \langle k \rangle_{BA} $. We set C as 0.8 when we calculate $S_{Sim}$.}
\label{fig3}
\end{figure*}

Following this, we observe the correlation between accuracy and the mean degree (Fig. \ref{fig3}).
Positive correlations appear in both metrics for all similarity indices.
This means that link prediction in general performs poorly in low mean degree, regardless of methodology.

\begin{figure*}
\includegraphics[width=13cm]{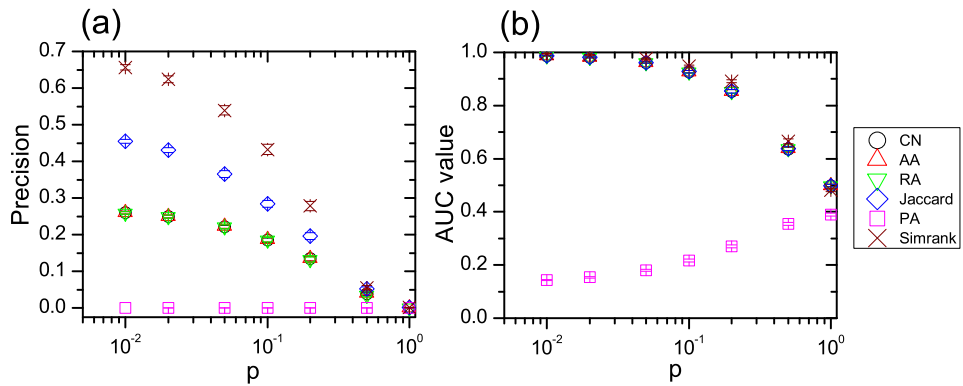}
\caption{Correlation between (a) precision and $p$ and (b) AUC value and $p$ in the WS model. We set C as 0.8 when we calculate $S_{Sim}$}
\label{fig4}
\end{figure*}

The correlation between accuracy and the rewiring probability in the WS model are then observed. (Fig. \ref{fig4}).
The precision of every index converges to near zero.
The WS model is equivalent to a random network at $p=1$.
Thus, the remaining network is also a random network because we randomly deleted the links.
Therefore, there is no structural gap between the original network and the remaining network, which means that the similarity indices fail to find some ``suspicious" node pairs.

The AUC value has a similar relation with precision.
When $p=1$, all indices except PA have an AUC value of 0.5, which is the same as the performance of a random classifier.
All parameters except PA show a negative correlation.

When we observe relative performance, clustering-based indices (CN, AA, RA and Jaccard) perform better in the WS model,
but PA exhibits the best performance in the BA model.
This difference might be rooted in the network structure and the attributes of the similarity indices.

\section{Discussion}

We tested link prediction using two network models: the WS model and the BA model.
Our results, indicate that some network features appear regardless of similarity indices and accuracy metrics.
First, a positive correlation between the mean degree and accuracy metrics is observed, which helps explain why a sparse network is difficult to predict.
A decrease in the number of unconnected links and the increase in topological information might contribute to an increase in both accuracy metrics.
Second, the AUC value is less dependent on $N$ than precision is.
In the BA model, PA is negatively correlated with $N$ whereas the AUC value is not correlated with it at all.
Other indices seems somewhat dependent on $N$ because the clustering coefficient of the BA model is negatively correlated with $N$.
Therefore, this correlation is not due to the size dependence of the AUC value, but that of network properties.
We can infer that the AUC value is a more appropriate measure of accuracy than precision for networks of different sizes.
Third, prediction accuracy on a random network ($p=1$ in the WS model) approaches the accuracy of a random classifier.
The randomness of the network for all indices approaches a random classifier in the statistical perspective, and thus the AUC values of all indices converge to 0.5.
Furthermore, the precision values of all indices converge to 0.006, which is identical to the precision value of a random classifier.

The different results for the similarity indices reflect a few of their intrinsic properties.
A statistically similar pattern is exhibited by the AUC values.
Clustering-based indices (CN, AA, RA and Jaccard) have almost the same AUC values for all conditions.
A similar result is also obtained for the precision values, but some difference exists.
The Jaccard index performs better than the other three indices (CN, AA and RA) in the WS model.
However, the precision of the Jaccard index in the BA model is lower than that of each of the three indices.
This difference might be based on the normalization property of the Jaccard index.
Furthermore, only PA exhibits a positive correlation with $p$ in the WS model.
Two accuracy metrics are negatively correlated with $p$ for all indices except PA.
PA performs better when the heterogeneity of degree is higher, and thus it shows a positive correlation with $p$.
Characteristics of the index reflect not only accuracy but also the correlation pattern of accuracy with different network parameters.

The numerical values of precision and the AUC value provide some evidence of more accurate results.
Clustering-based indices record very high precision and the AUC values in the WS model because the model is well-clustered,
whereas PA performs more poorly than a random classifier.
The degree of any two nodes decreases when we delete the link between them.
Since the degree distribution of the WS model is concentrated around the mean degree, PA of the missing links is usually lower than that of any other pair.
On the contrary, in the BA model, PA performs relatively well.
The BA model follows power-law degree distribution and is not well clustered, so PA shows better performance than in the WS model.
However, its precision and the AUC value are quite low.
PA has high value when each node in a pair has a high degree, but such a pair can be unconnected in the BA model because hub nodes in the BA model are usually connected with low-degree nodes.

Our study for model networks has revealed several relationships between network parameters and accuracy metrics.
However, the model networks employed in this study do not reflect all features of real-world networks.
More practical models need to be considered to better understand the characteristics of link prediction in empirical networks, and a large number of indices need to be taken into account in these models.

\section{Conclusions}

We tested link prediction using two basic network models and various indices to observe the statistical relationship between network parameters and accuracy of link prediction.
We applied six similarity indices (CN, AA, RA, Jaccard, PA and Simrank), and used precision and the AUC value to measure accuracy.
We observed a statistical relationship between accuracy and network parameters.
Some features, such as a positive correlation with the mean degree or the size independence of AUC values, indicate that all similarity indices and both accuracy metrics are positively correlated with the mean degree in both models, and that the AUC value is a more size-independent metric than precision.
Other relationship shows some intrinsic properties of similarity indices.
Clustering-based indices exhibit similar patterns and values for every case, but PA differs from other indices.
This difference is based on the intrinsic property of PA, whereby PA depends on the degree and other indices depend on clustered structure of a network.
This framework is expected to contribute to a better understanding of how link prediction methodology works even for more complex cases.

This work was supported by Mid-career Researcher Program through the National Rresearch Foundation of Korea (NRF) grant funded by the Korea government (MSIP) (NRF-2013-R1A2A2A04017095) and Basic Science Research Program through the National Research Foundation of Korea (NRF) funded by the Ministry of Education, Science and Technology (NRF-2010-0021987).

\section*{References}
\biboptions{sort&compress}
\bibliographystyle{apsrev}
\bibliography{accuracy_test_for_link_prediction}
\end{document}